\documentclass[fp,twocolumn]{jpsj3}
\usepackage{txfonts}
\usepackage{amsmath}%
\usepackage{bm}%
\usepackage{graphicx}
\usepackage{color}

\def\ea4{EuAl$_4$}
\def\eg4{EuGa$_4$}
\def\ex4{Eu$X_4$}
\def\tcdw{$T_{\rm CDW}$}
\def\tn{$T_{\rm N}$}
\def\tna{$T_{\rm N1}$}
\def\tnb{$T_{\rm N2}$}
\def\tnc{$T_{\rm N3}$} 
\def\tnd{$T_{\rm N4}$}
\def\da{${\delta}_1$}
\def\dpa{${\delta}_2$}
\def\dc{${\delta}_c$}
\def\dd{(${\delta}_2~{\delta}_2~0$)}
\def\d00{(${\delta}_1~0~0$)}

\def\qvec{${\bm q}$}
\def\qveca{${\bm q}_{1}$}
\def\qvecab{${\bm q}_1^{\prime}$}
\def\qvecb{${\bm q}_{2}$}
\def\qvecc{${\bm q}_{\rm CDW}$}
\def\qvecaf{${\bm q}_{1}$~=~{\d00}}
\def\qvecbf{${\bm q}_{2}$~=~{\dd}}
\def\qveccf{${\bm q}_{\rm CDW}$~=~(0~0~{\dc})}

\newcommand{\bvec}[1]{\mbox{\boldmath $#1$}}

\title{Charge-Density-Wave Order and Multiple Magnetic Transitions in Divalent Europium Compound EuAl$_4$}

\author{Koji Kaneko$^{1,2}$\thanks{koji.kaneko@j-parc.jp}, Takuro Kawasaki$^2$, Ai Nakamura$^{3,4}$, Koji Munakata$^5$, Akiko Nakao$^5$,\\ Takayasu Hanashima$^5$, Ryoji Kiyanagi$^2$, Takashi Ohhara$^2$,  
Masato Hedo$^6$, Takao Nakama$^6$, and\\ Yoshichika \=Onuki$^{6,7}$}
\inst{$^1$Materials Sciences Research Center, Japan Atomic Energy Agency, Tokai, Naka, Ibaraki 319-1195, Japan\\
$^2$J-PARC Center, Japan Atomic Energy Agency, Tokai, Naka, Ibaraki 319-1195, Japan\\
$^3$Graduate School of Engineering and Science, University of the Ryukyus, Nishihara, Okinawa 903-0213, Japan\\
$^4$Institute for Materials Research, Tohoku University, Oarai, Ibaraki 311-1313, Japan\\
$^5$Comprehensive Research Organization for Science and Society, Tokai, Ibaraki 319-1106, Japan\\
$^6$Faculty of Science, University of the Ryukyus, Nishihara, Okinawa 903-0213, Japan\\
$^7$RIKEN Center for Emergent Matter Science, Wako, Saitama 351-0198, Japan\\
} 

\abst{
Multiple transition phenomena in divalent Eu compound {\ea4} with the tetragonal structure were investigated via the single-crystal time-of-flight neutron Laue technique. 
At 30.0~K below a charge-density-wave (CDW) transition temperature of {\tcdw}~=~140~K, superlattice peaks emerge near nuclear Bragg peaks described by an ordering vector {\qvecc}~=~(0~0~{\dc}) with {\dc}~${\sim}$~0.19.
In contrast, magnetic peaks appear at {\qvecbf} with {\dpa}~=~0.085 in a magnetic-ordered phase at 13.5~K below {\tna}~=~15.4~K.
By further cooling to below {\tnc}~=~12.2~K, the magnetic ordering vector changes into {\qvecaf} with {\da}~=~0.17 at 11.5~K and slightly shifts to {\da}~=~0.194 at 4.3~K.
No distinct change in the magnetic Bragg peak was detected at {\tnb}=13.2~K and {\tnd}=10.0~K.
The structural modulation below {\tcdw} with {\qvecc} is characterized by the absence of the superlattice peak in the (0~0~$l$) axis.
As a similar CDW transition was observed in SrAl$_4$, the structural modulation with {\qvecc} could be mainly ascribed to the displacement of Al ions within the tetragonal $ab$-plane.
Complex magnetic transitions are in stark contrast to a simple collinear magnetic structure in isovalent {\eg4}.
This could stem from different electronic structures with the CDW transition between two compounds.
}

\begin{document}
\maketitle

\section{Introduction}
The interplay of exchange interactions and single-ion anisotropy are key ingredients for realizing diverse magnetic properties of rare-earth compounds. 
In most of the rare-earth ions, magnetism is carried by a large orbital moment and is therefore affected by single-ion anisotropy via the crystalline electric field.
Eu$^{2+}$ and Gd$^{3+}$ have a unique state without the orbital angular momentum, namely, $J=S=7/2$ ($L$=0).
It provides an ideal opportunity to study collective spin behavior without single-ion anisotropy.
In addition, recent studies have revealed that characteristic topological spin textures, such as a magnetic skyrmion lattice, are formed in Eu- and Gd-based pure spin systems.
In EuPtSi with a cubic chiral structure having the same space group as the prototypical skyrmion MnSi,
the formation of the skyrmion lattice with a short periodicity of 18~{\AA} is characterized by a huge topological Hall effect and a triple-{\qvec} magnetic order.\cite{Kakihana2018,Kaneko2019,Takeuchi2019,Tabata2019,Sakakibara2019,Kakihana2019,Homma2019}
In Gd-based compounds, skyrmion lattices with the similar short periodicity are formed even in the centrosymmetric structure.\cite{Kurumaji2019,Hirschberger2019} 
These discoveries engender intensive research on Eu- and Gd-based compounds. 

As a divalent Eu system, EuGa$_4$ and EuAl$_4$ offer an interesting possibility to investigate 4$f$-based spin magnetism.
Both compounds crystallize in the tetragonal BaAl$_4$-type structure with the space group $I4/mmm$.\cite{DeVries1985,Bobev2004,Nakamura2013,Nakamura2015a}
This structure can be regarded as a parent variant of Eu$T_2X_2$, where both $T$ and $X$ sites are occupied by Ga/Al.
Eu ions in both compounds are reported to have a divalent state; they are considered as a pure spin system with $J=S=7/2$.
Indeed, the isotropic magnetic susceptibilities in the paramagnetic state were revealed in both compounds.
However, these isovalent compounds show contrasting behavior at low temperatures.
{\eg4} orders antiferromagnetically below {\tn}=16~K.\cite{Nakamura2013}
The antiferromagnetic order can manifest as a cusp in the magnetic susceptibility along the magnetic easy $a$-axis.
The specific heat shows a sharp ${\lambda}$-type anomaly at {\tn}. 
A single-crystal neutron diffraction experiment revealed that the antiferromagnetic structure can be described by a simple ordering vector with {\bvec{q}}=(0~0~0), in which
the ordered moment of Eu reaches 7~${\mu}_{\rm B}$ as expected for $S$=7/2 and orients parallel to the $a$-axis.\cite{Kawasaki2016}
The almost isotropic {\bvec{B}}-$T$ phase diagram is reproduced by assuming $S$=7/2 as well.
These findings are consistent with those of microscopic studies by NMR and M\"ossbauer spectroscopies.\cite{NIKI2014,Yogi2013,Homma2014}

\begin{figure*}[ttttt]
	\centering
	\includegraphics[width=16cm]{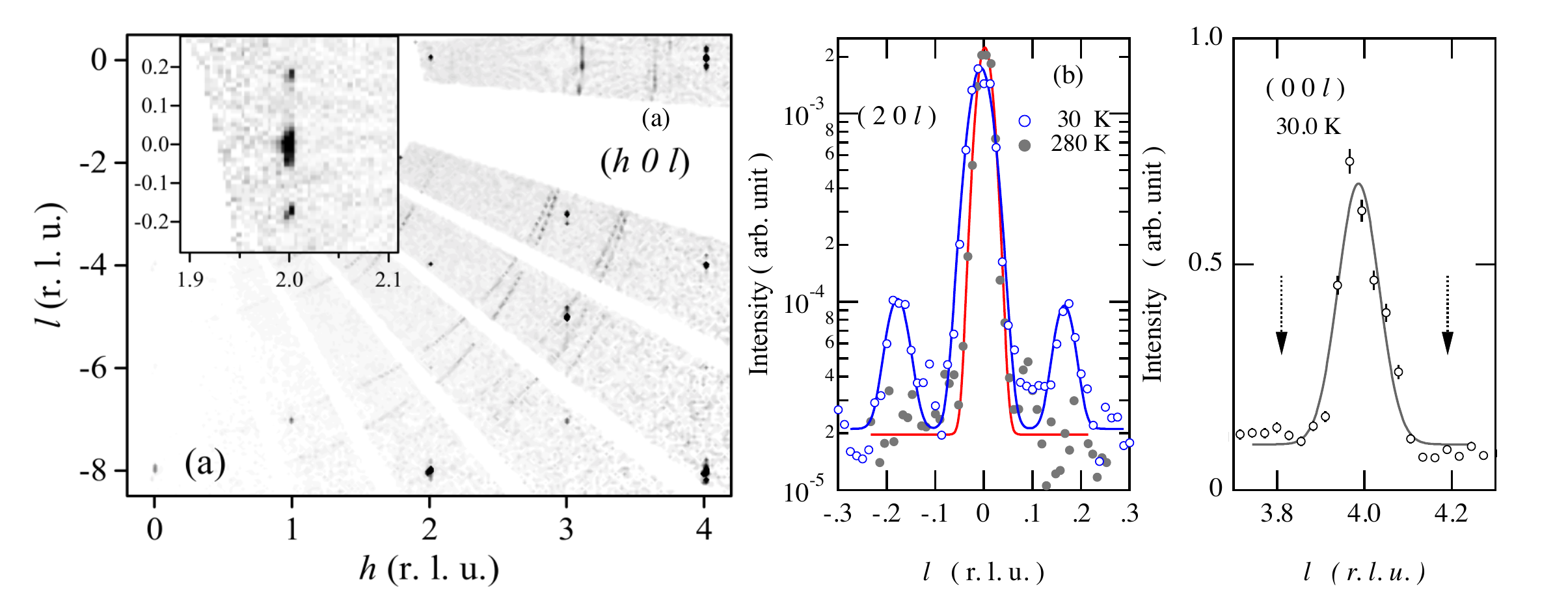}
	\vspace{5mm}
	\caption{(Color online) (a) Neutron scattering intensity map of the reciprocal ($h~0~l$) plane of {\ea4} at 30.0~K. Enlarged map around (2~0~0) is shown as an inset. 
	(b,c) Cut along (0~0~$l$) across the nuclear peak at (b) \bvec{Q}=(2~0~0) at 30.0~K ($T<${\tcdw}) and 300~K ({\tcdw}$<T$), and (c) across \bvec{Q}=(0~0~6) at 30.0~K.}
	\label{cdwmap}
\end{figure*}
							
On the other hand, {\ea4} exhibits rather complex physical properties; it undergoes five successive transitions upon cooling from room temperature to 4~K.\cite{NAKAMURA2014,Nakamura2015a}
The first transition at 140~K is observed as a hump in electrical resistivity and thermopower measurements, and is ascribed to the charge-density-wave (CDW) order.
Indeed, a decrease in carrier density below this transition was revealed by both the magnetoresistance and Hall effect.\cite{Araki2014}
The second-order magnetic transition occurs at {\tna}=15.4~K, which is followed by the remaining three successive magnetic transitions in a narrow temperature range, namely at {\tnb}=13.2~K, {\tnc}=12.2~K, and {\tnd}=10.0~K. 
The magnetic susceptibility shows several kinks at the corresponding transition temperatures along both the $ab$-plane and $c$-axis.
The consecutive magnetic transitions in {\ea4} are also confirmed in the magnetization curve, where four-step magnetic transitions were observed along the $c$-axis at 2~K.  
The specific heat exhibits a distinct ${\lambda}$-type anomaly at {\tna} and an abrupt jump at {\tnc}, indicating the second- and first-order nature of the transition, respectively,  whereas anomalies at {\tnb} and {\tnd} were small in magnitude.  
In contrast to the similar temperature scale of {\tn} between {\eg4} and {\ea4}, the complex transition sequence and the presence of CDW order are unique characteristics of {\ea4}.
Moreover, recent studies suggest the emergence of the topological Hall effect in {\ea4}, which increases interest on its fundamental properties\cite{Shang2021}.
Here, single-crystal neutron scattering experiments were carried out to obtain insights into transition phenomena in {\ea4}.

\section{Experiment}
Single crystals of {\ea4} were grown by the Al self-flux method using naturally occurring europium.
Details of sample growth have been published elsewhere.\cite{Nakamura2015a,Nakamura2013}
The sample was cut to a rectangular shape with dimensions of roughly $2.0~{\times}~1.7~{\times}~0.6$~mm$^3$. 

Single-crystal neutron diffraction experiments on {\ea4} were carried out on the time-of-flight neutron Laue diffractometer SENJU installed at BL18 of the Materials and Life Science Experimental Facility (MLF) of J-PARC in Tokai, Japan.\cite{Ohhara2016}
The first frame, which corresponds to shorter wavelengths ranging from 0.4 to 4.0~{\AA}, was employed to cover a wide reciprocal lattice space and to reduce the strong neutron absorption of Eu.
The sample was glued to a vanadium rod and mounted onto a piezo-motor-based two-axes goniometer.\cite{Ohhara2015}
The goniometer was directly attached to a cold finger of a closed-cycle refrigerator that can be cooled down to 4~K.
To regulate the temperature precisely, a thermometer was mounted on the top of the goniometer close to the sample.
Representative diffraction data were collected at 4.3~K ($T<${\tnd}), 30.0~K ({\tna}$<T<${\tcdw}), and 300 K ({\tcdw}$< T$) at several orientations for 4 h.
Successive magnetic transitions were investigated with a shorter counting time of 2 h at a fixed orientation from 4.3 to 30.0~K. 
Data reductions and the visualization of the diffraction intensity in reciprocal lattice space were performed using the software STARGazer~\cite{Ohhara2009}.

\section{Results}
Figure~{\ref{cdwmap}}(a) shows the neutron scattering intensity in a part of the ($h~0~l$) reciprocal lattice plane at 30.0~K below {\tcdw}. 
A large number of nuclear Bragg reflections were clearly observed despite the strong neutron absorption of Eu.
The refined lattice constants at room temperature are in good agreement with those in a previous report\cite{Nakamura2015a}, with $a$=4.404(1)~{\AA} and $c$=11.148(2)~{\AA}.
Nuclear Bragg peaks exist at $h+l=even$, which satisfies the extinction rule for the body-centered $I4/mmm$.

\begin{figure}
	\centering
	\includegraphics[width=7cm]{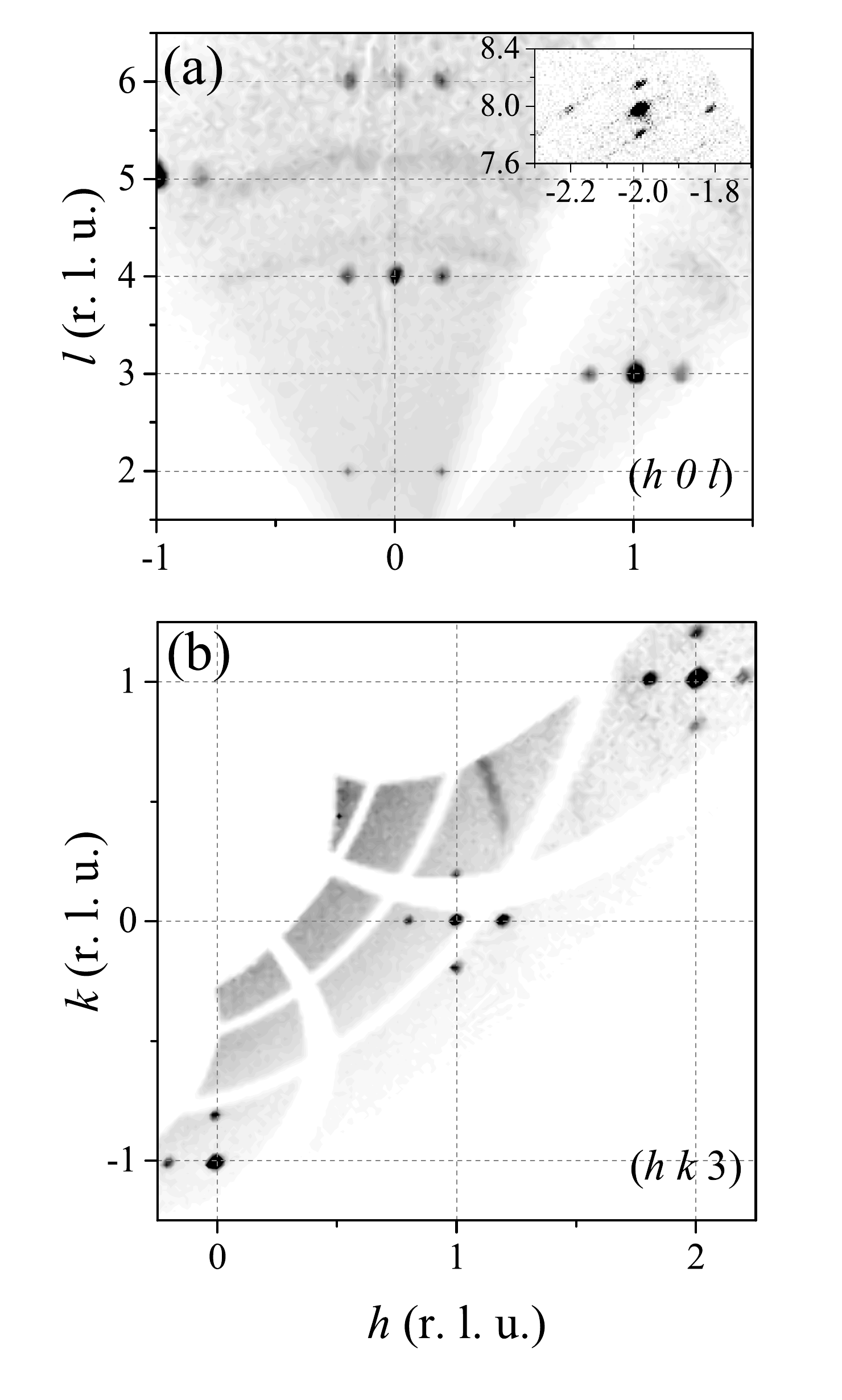}
	\caption{Neutron scattering intensity map of the reciprocal (a) ($h~0~l$) and ($h~k~3$) planes recorded at 4.3~K in the ground state ($T<${\tnd}) of {\ea4}.}
	\label{magmap1}
\end{figure}

Moreover, additional superlattice peaks are found close to the nuclear Bragg peaks. 
The inset shows the magnified intensity map at around (2~0~0).
Superlattice peaks are observed as satellite peaks with splitting along the (0~0~$l$) axis.
\begin{figure*}[ttt]
	\begin{center}
		\includegraphics[width=16cm]{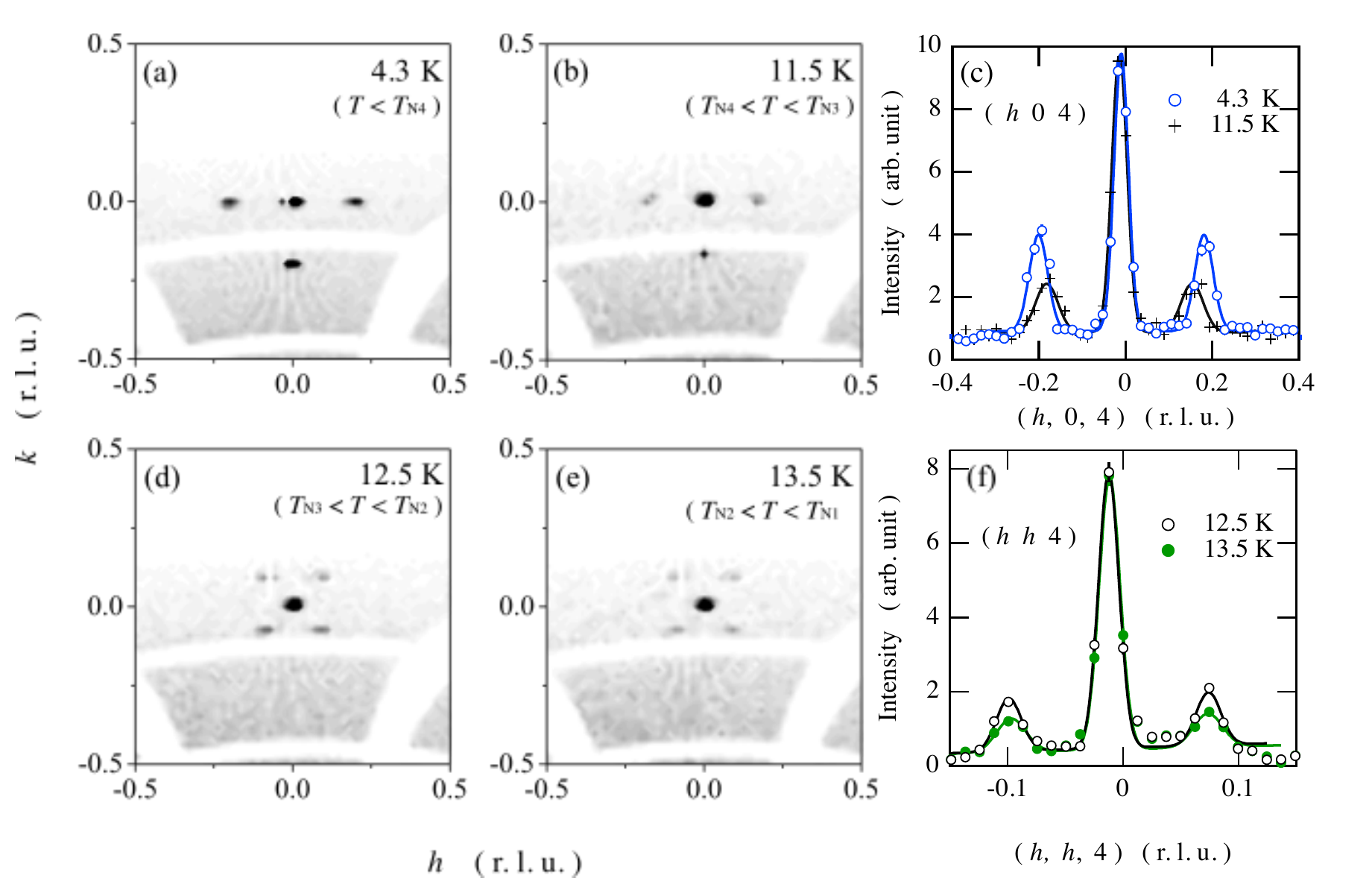}
		\caption{(Color online) Neutron scattering intensity maps of the reciprocal ($h~k~4$) plane at each magnetic ordered phase, (a)~4.3~K ($T<${\tnd}), (b)~11.5~K ({\tnd}$<T<${\tnc}), (d)~12.5~K ({\tnc}$<T<${\tnb}), and (e)~13.5~K ({\tnb}$<T<${\tna}) of {\ea4}.
		 Line profile along (c) ($h$~0~0) direction for 4.3 and 11.5~K and (f) ($h~h,$~0) direction for 12.5 and 13.5~K through (0~0~4).}
		\label{magmap2}
	\end{center}
\end{figure*}
Figure~{\ref{cdwmap}}(b) shows a cross-sectional profile along the (0~0~$l$) axis through the 2~0~0 Bragg peak measured below and above {\tcdw}, plotted in the logarithmic scale.
Satellite peaks appear below {\tcdw} and have an integrated intensity of roughly 1/30 as compared with the 2~0~0 nuclear peak,
whereas the intensity of the nuclear Bragg peak slightly decreases in magnitude below {\tcdw}.
A simple Gaussian fitting with symmetric position with respect to a nuclear peak gives satellite peak positions as $l={\pm}0.19(1)$, namely, the propagation vector for the superlattice peak in this phase is {\qveccf} with {\dc}${\sim}$0.19.
No higher-order harmonics of the satellite were observed at 2{\dc} and 3{\dc} within the present accuracy.

In terms of intensity, superlattice peaks become apparent at higher ${\bm Q}$, as seen at around 4~0~8, indicating its structural origin.
Comparison of the intensities of satellites in a similar ${\bm Q}$ range, such as (2~0~8), (3~0~5), and (4~0~0), reveals that the peak intensity tends to increase with increasing in-plane component increases, namely, with decreasing $l$.
Indeed, satellite peaks are absent in the (0~0~$l$) line, as demonstrated at around (0~0~4) in Fig.~{\ref{cdwmap}}(c).

The neutron diffraction patterns in the ground state of {\ea4} are summarized in Fig.~\ref{magmap1}.
Intensities at a low-$Q$ part of the ($h~0~l$) and ($h~k~3$) reciprocal lattice planes recorded at 4.3~K are shown in Figs.~\ref{magmap1}(a) and \ref{magmap1}(b), respectively.
In Fig.~\ref{magmap1}(a), nuclear Bragg peaks at (0~0~2), (0~0~4), (0~0~6), and (1~0~3) accompany additional satellite peaks on both sides along the ($h$~0~0) line. 
In contrast to the satellites in the CDW phase, the existence of clear satellite peaks at the low-${\bm Q}$ region indicates its magnetic origin.
In addition, the satellite peaks were clearly observed near the (0~0~$l$) line,  as represented near (0~0~6).
Because the magnetic neutron scattering intensity is proportional to the square of the moment perpendicular to the scattering vector ${\bm Q}$,
the result indicates that the moment does not orient parallel to the $c$-axis, and at least has an in-plane component.

The inset of Fig.~\ref{magmap1}(a) displays the ($h~0~l$) intensity map at around (2~0~8).
The superstructural peaks with {\qvecc} and the magnetic peaks with {\qveca} were observed along the $l$ and $h$ lines, respectively.
The superlattice peaks along the $l$ direction have a modulation close to {\qveccf} of {\dc}=0.18(1) and do not exist in the low-$Q$ region.
This observation confirms that the structural CDW peak coexists with the magnetic peak in the ground state. 
In the intensity data of the ($h~k~3$) plane in Fig.~\ref{magmap1}(b), superlattice peaks were observed at (1${\pm}$0.19~0~3) and (1~${\pm}$0.19~3) and its equivalent positions at around (0~${\bar 1}$~3) and (2~1~3), respectively.
The result provides evidence that the magnetic order in the ground state is described by a propagation vector with {\qvecaf} and equivalent {\qvecab}=(0~{\da}~0).
A fitting of multiple peaks gives {\da}=0.194(5) at 4.3~K.
No higher-order harmonics of {\qvecaf} were observed. 
The absence of a magnetic peak at the commensurate position, such as (0~0~3), manifests that the magnetic structure of {\ea4} is different from that of {\eg4} with {\qvec}=(0~0~0).

Concerning consecutive magnetic transitions, Fig.~\ref{magmap2} illustrates the ($h~k~0$) neutron intensity map at around (0~0~4) through the four magnetic transitions from {\tna} to {\tnd}, measured upon heating.
At 4.3~K in the ground state, magnetic peaks locate at (${\pm}${\da}~0~4) and (0~${\pm}${\da}~4).
By heating up to 11.5~K across {\tnd}, the magnetic peak holds the same symmetry as the ground state as plotted in Fig.~\ref{magmap2}(b).
On the other hand, the periodicity {\da} slightly changes in this temperature range.
Figure~\ref{magmap2}(c) shows the line profile along the $h$-axis through (0~0~4).
The fitting with the symmetrically positioned Gaussian against nuclear peaks over several $Q$ positions results in a change in {\da} from 0.194(5) at 4.3~K to 0.17(1) at 11.5~K. 

A marked change in the magnetic peak position was detected upon heating up to 12.5 K across {\tnc} as shown in Fig.~\ref{magmap2}(d); 
magnetic peaks move from (${\pm}${\da}~0~4) and (0~${\pm}${\da}~4) to the diagonal positions (${\pm}${\dpa}~${\pm}${\dpa}~4) and (${\pm}${\dpa}~${\mp}${\dpa}~4) with {\dpa}${\sim}$0.085 at 12.5~K.
This result reveals that the magnetic ordering vector changes from {\qvecaf} to {\qvecbf} in the first-order transition at {\tnc}.
No higher-order harmonics were observed for the ordering vector {\qvecb} as well.
Upon passing through {\tnb}, the magnetic peak with {\qvecbf} remains at the same position, and no additional change was observed (see Fig.~\ref{magmap2}(e)).
To see this in detail, the profile along the ($h~h$~0) line through (0~0~4) is plotted in  Fig.~\ref{magmap2}(f).
A similar fitting as in Fig.~\ref{magmap2}(c) along the ($h~h$~0) direction gives {\dpa} to be 0.085(4) and 0.086(4) at 12.5 and 13.5~K, respectively.
Namely, no distinct shift of the magnetic peak was observed within the present accuracy. 
By further heating above {\tna}, the disappearance of the magnetic peak was confirmed. 
Throughout the consecutive magnetic transitions, no remarkable variation was also observed in the nuclear Bragg peak at (0~0~4).

\section{Discussion}
Concerning the CDW transition, the present neutron diffraction study reveals that the transition at {\tcdw} is characterized by structural satellite peaks with the incommensurate ordering vector {\qveccf} with {\dc}${\sim}$0.19.
This is consistent with a recent single-crystal X-ray diffraction study that also revealed the appearance of superlattice peaks with {\qvecc}.\cite{Shimomura2019}
One of the characteristics of the structural satellite peaks with {\qvecc} is their absence along the (0~0~$l$) line and stronger peaks with increasing the in-plane component, namely, decreasing $l$.
This feature suggests that this structural modulation arises from a displacement within the $ab$-plane. 
In {\ea4}, there are three crystallographic sites, namely, the 2$e$ site for Eu and the $4d$ and $4e$ sites for Al.
Whereas the $4e$ site has an extra reflection condition for $hkl$ with $l$=2$n$, the observed superlattice peaks do not have any additional extinction.
In other words, the result does not provide supplementary information to identify which ions are responsible for the displacement.  
Note that a similar CDW transition was found in the isostructural compound SrAl$_4$ without the 4$f$ element\cite{Nakamura2016,Shimomura2019,Niki2020,Niki2015a}.
It implies that the Eu ion is not involved in the displacement, and the Al ion could be responsible for the displacement.
Indeed, the electronic structure of {\ea4} disclosed by de Haas van Alphen measurement, photoelectron spectroscopy, and band calculations 
shows close similarity to SrAl$_4$, where 4$f$ electrons lie well below the Fermi level\cite{Nakamura2015a,Kobata2016,Nakamura2016}.
These findings support that the CDW transition is driven by the displacement of the Al ion.
As for its origin, one possible scenario is the nesting of the Fermi surface.
However, the reported electronic structure indicates that the topology of the Fermi surface does not have clear instability at {\qveccf}.
In addition, the photoelectron spectroscopy did not show a notable change across {\tcdw}.
Although the origin of CDW remains unknown at present, it is revealed that there is a clear discrepancy in the Fermi surface between {\eg4} and {\ea4}.

In addition to the CDW transition, the distinct differences between {\ea4} and {\eg4} are the multiple magnetic transitions and the complex magnetic ordering vector.
Throughout consecutive magnetic transitions, the magnetic ordering vectors of {\ea4} have the in-plane components in {\qvecaf} and {\qvecbf}, 
which are dissimilar to the structural one with the component along $c^*$, and a simple commensurate structure with {\qvec}=(0~0~0) in {\eg4}.\cite{Kawasaki2016}
A contrasting magnetism in isovalent compounds is also revealed in other spin systems.
EuAs$_3$ shows two magnetic transitions, namely, an incommensurate sinusoidal intermediate phase and a commensurate ground state.\cite{Chattopadhyay1986}
On the other hand, isovalent substitution of As by P leads to a change of the ground state into a spiral phase with an incommensurate structure, although an intermediate phase with the incommensurate sinusoidal structure is common to EuAs$_3$.\cite{Chattopadhyay1987}
In the BaAl$_4$-derived structural family, diverse magnetic ground states are realized in Gd$T_2$X$_2$, where $T$ is transition metal elements and $X$=Si and Ge.\cite{Mallik1998,Good2005,Barandiaran1988}
Several mechanisms for a variety of magnetic ground states and successive transitions are proposed in terms of anisotropic and/or competing exchange interactions as well as high-order contributions.\cite{Kobler1999,Kobler1998}
A fundamental magnetic interaction in {\ex4}($X$=Al and Ga) is the Runderman--Kittel--Kasuya--Yoshida interaction based on the oscillatory exchange interaction between localized 4$f$ electrons via conduction electrons.
The CDW transition takes place only in {\ea4}, which results in/from the change in the Fermi surface.
This difference between two compounds could lead to contrasting magnetic interactions.

Quadruple magnetic transitions in {\ea4} are quite unique among Eu- and Gd-based spin systems.
This study revealed that the transition at {\tna} and {\tnc} corresponds to the incommensurate order with {\qvecaf} and the order--order transition to {\qvecbf}, respectively. 
On the other hand, neither the anomaly in the magnetic Bragg peaks with {\qveca} and {\qvecb} nor the appearance/disappearance of higher-order harmonics was found at {\tnb} and {\tnd} within the present accuracy.
In other words, the nature of the transition at {\tnb} and {\tnd} remains unclarified.
There are several possibilities to account for these transitions.  
One is a transition from an amplitude-modulated-type to an equal-moment-type structure, observed in several Eu and Gd compounds.
This is characterized by the appearance of higher-order satellite peaks, which are usually quite weak.
In the isotropic system without single-ion anisotropy, applicable to a pure spin system, the mean-field analysis on the specific heat provides a clue to the type of magnetic order.\cite{Blanco1991a,Rotter2001}
A specific heat jump at the magnetic transition to an equal-moment order is a factor of 1.5 larger than those to a collinear amplitude-modulated magnetic structure.
The specific heat jump at {\tna} in {\ea4} is significantly reduced from that in {\eg4}, although the multifold transitions in {\ea4} make it difficult to analyze the data.
This finding implies that a simple equal-moment magnetic structure is not formed at {\tna}.
The other possibility is the lock-in transition from incommensurate to commensurate structures. 
As for the transition at {\tnb}, the modulation {\dpa} in {\qvecbf} is derived as 0.085(4) and 0.086(4) for 12.5 and 13.5~K, respectively.
Namely, no visible change in {\dpa} was found beyond the experimental accuracy, whereas it is not far from 0.083${\simeq}$1/12.
In contrast, a clear peak shift was observed between 4.3 and 11.5~K through {\tnd}, as displayed in \ref{magmap2}(c).
The modulation {\da} in {\qvecaf} is changed from 0.17(1) at 11.5~K to 0.194(5) at 4.3~K.
The value at 4.3~K is slightly less than 0.2=1/5, but is quite close.
To examine this scenario, determination of the detailed temperature dependence of the peak position will be useful and currently in progress.

Concerning the magnetic structure in the ground state, two possible models for {\qvecaf} exist: a single-{\qvec} structure with a multiple-domain state, or a double-{\qvec} structure with a single-domain state. 
Although the present result does not allow us to distinguish between two models, a high-resolution single-crystal X-ray diffraction study could provide a supplementary input;
a slight distortion from tetragonal to lower symmetry, such as an orthorhombic, takes place below {\tnc}.\cite{Shimomura2019}
This distortion is minute, as the ratio $b/a$ in the orthorhombic case is roughly 0.998, identified by the satellite peak at (6.008~0~0).
Although the direction along the ($h$~0~0) corresponds to the time-of-flight axis in SENJU, which has higher resolution, sensitivity of ${\Delta}d/d{\sim}$0.5\% is insufficient to detect the reported minute distortion.    
Even if the distortion is small, the existence of the in-plane distortion can lift the degeneracy of {\qvecaf} and equivalent {\qvecab}=(0~{\da}~0), and therefore could favor the single-{\qvec} state as the ground state.
This coupling between magnetism and lattice distortion could be the origin of the complex multistep transition in {\ea4}.
In addition, a possible canting of the magnetic moment from the $c$-axis suggested by recent NMR and M\"ossbauer spectroscopies\cite{Niki2020} can bring another degree-of-freedom.
The present data do not allow the corresponding magnetic structure to be revealed, as reliable quantitative analysis is difficult owing to strong wavelength-dependent neutron absorption.
Further neutron diffraction experiments with quantitative analysis will be necessary to reveal the magnetic structure. 

\section{Summary}
Single-crystal neutron diffraction experiments were performed on {\ea4} using the time-of-flight Laue technique. 
Superlattice peaks described by {\qvecc} were observed at 30.0~K below the CDW transition {\tcdw}, which is characterized by the absence of the peak along the (0~0~$l$) axis. 
Although these superlattice peaks survive down to the base temperature of 4.3~K, magnetic peaks were observed at {\qvecaf} with {\da}=0.194. 
Heating above {\tnd} leads to a slight peak shift to {\da}=0.17 at 11.5~K while keeping {\qvecaf}.
A marked change takes place on heating above {\tnc}, where the magnetic ordering vector changes from {\qvecaf} to {\qvecbf} with {\dpa}=0.085.
Complex incommensurate magnetic transitions in {\ea4} are in clear contrast to a simple collinear commensurate structure in isovalent {\eg4}.

\begin{acknowledgment}
We would like to thank M. Yogi, H. Niki, Y. Tokunaga, Y. Homma, S.-i. Fujimori, and H. Yamagami for valuable discussions. 
We also thank N. Aso for collaboration in the neutron scattering experiment on the related compound {\eg4}.
This work was partly supported by Grants-in-Aid for Scientific Research (C) (Nos. 24540336, 26400348, and 16K05031) and (B) (No. 20H01864) from the Japan Society for the Promotion of Science.
The neutron diffraction experiment on SENJU at MLF J-PARC was performed under a user program (Proposal No. 2014A0080).

\end{acknowledgment}

\bibliographystyle{./jpsj}

\end{document}